\documentclass[twocolumn,showpacs,superscriptaddress,nofootinbib]{revtex4-1}
\usepackage[T1]{fontenc} 
\usepackage{amsmath}
\usepackage{amssymb}
\usepackage{amsfonts}
\usepackage{graphicx}
\usepackage{enumitem}
\usepackage{bm}
\usepackage{rotating}
\usepackage{hyperref}
\usepackage{xcolor}
\usepackage{array}
\usepackage{lmodern}
\usepackage[normalem]{ulem}
\usepackage{braket}
\usepackage{tensor}
\usepackage{mathrsfs}
\usepackage{float}

\newcommand{\be}{\begin{equation}}
\newcommand{\ee}{\end{equation}}
\newcommand{\ba}{\begin{eqnarray}}
\newcommand{\ea}{\end{eqnarray}}
\newcommand{\beq}{\begin{equation}}
\newcommand{\eeq}{\end{equation}}
\newcommand{\beqa}{\begin{eqnarray}}
\newcommand{\eeqa}{\end{eqnarray}}
\newcommand{\nn}{\nonumber}
\newcommand{\sx}{\mathsf{x}}

\usepackage[normalem]{ulem}

\begin{document}

\title{Quantum Detection of Inertial Frame Dragging}

\author{Wan Cong}
\email{wcong@uwaterloo.ca}
\address{Department of Physics and Astronomy, University of Waterloo, Waterloo,
Ontario, Canada, N2L 3G1}
\address{Perimeter Institute, 31 Caroline St., Waterloo, Ontario, N2L 2Y5, Canada}

\author{Ji\v r\'i Bi\v c\'ak}
\email{bicak@mbox.troja.mff.cuni.cz }
\address{Institute of Theoretical Physics, Faculty of Mathematics and Physics, V Hole\v sovi\v ck\'ach 2, 180 00 Prague 8, Czech Republic}

\author{David Kubiz\v n\'ak}
\email{dkubiznak@perimeterinstitute.ca}
\address{Perimeter Institute, 31 Caroline St., Waterloo, Ontario, N2L 2Y5, Canada}
\address{Department of Physics and Astronomy, University of Waterloo, Waterloo,
Ontario, Canada, N2L 3G1}

\author{Robert B. Mann}
\email{rbmann@uwaterloo.ca}
\address{Department of Physics and Astronomy, University of Waterloo, Waterloo,
Ontario, Canada, N2L 3G1}
\address{Perimeter Institute, 31 Caroline St., Waterloo, Ontario, N2L 2Y5, Canada}

\begin{abstract}
A relativistic theory of gravity like general relativity produces phenomena differing fundamentally from Newton's theory. An example, analogous to electromagnetic induction, is gravitomagnetism, or the dragging of inertial frames by mass-energy currents. These effects have recently been confirmed by classical observations. Here we show, for the first time, that they can be observed by a quantum detector. We study the response function of Unruh De-Witt detectors placed in a slowly rotating shell. We show that the response function picks up the presence of rotation even though the spacetime inside the shell is flat and the detector is locally inertial. The detector can distinguish between the static situation when the shell is non-rotating and the stationary case when the shell rotates and the dragging of inertial frames, i.e. gravitomagnetic effects, arise. Moreover, it can do so when the detector is switched on for a finite time interval within which a light signal cannot travel to the shell and back to convey the presence of rotation.
\end{abstract}

\maketitle

\section{Introduction}

Frame-dragging, also known as the Lense--Thirring effect  \cite{Lense:1918zz, thirring1918}, is a general-relativistic effect that arises due to moving, in particular rotating, matter \cite {CiuW} and ``rotating'' gravitational waves \cite{BiLe,BaLe}.  If a gyroscope is located in the vicinity of a rotating body, it will keep its direction with respect to the axes of a local inertial frame at the same place but both the inertial axes and the gyroscope will be rotating with respect to static distant observers (``fixed stars'' at asymptotically flat infinity). Its  profound explicit manifestation can be seen for a rotating black hole, which drags particles into co-rotation, the dragging becoming so strong inside the ergosphere that no particle there can remain at rest with respect to fixed stars \cite{MTW}. Frame-dragging is also behind various astrophysical phenomena such as relativistic jets and the Bardeen-Petterson effect~\cite{Bardeen:1975zz}, which aligns accretion disks perpendicular to the axis of a rotating black hole.

In addition, frame-dragging \textit{inside} a rotating shell was taken by Einstein to be in support of Mach's principle. For a nice discussion on Mach's principle, dragging effects and their impact on astrophysics and cosmology, see \cite{Barbour:1995iu} (also \cite{Pfister:2015ftd, CiuW}). Consider a slowly rotating material shell \cite{Pfi,PhysRev.143.1011}. Observers inside the shell who are at rest with respect to distant fixed stars will find that a particle moving inside the shell experiences a Coriolis acceleration (the centrifugal acceleration is of the second order in the shell's angular velocity). These observers are \textit{not inertial}, therefore fictitious forces arise. 

For \textit{inertial observers}, without looking at or outside the rotating shell, there is no way of determining, by employing classical physics,  whether they are surrounded by a rotating shell. They can in principle determine its rotation by, for example, sending out a spherical pulse which, upon
reflection, will experience a differential Doppler effect, with different shell latitudes Doppler shifting differently. Meanwhile, frame-dragging outside a rotating body, the Earth, has taken the Gravity Probe B satellite mission \cite{GPB,Everitt_2015} almost a half-century since its inception to detect.

In this paper, we show for the first time that frame-dragging inside a slowly rotating shell can be observed by an inertial quantum Unruh-DeWitt (UDW) detector \cite{DeWitt:1980hx}. Specifically, it can do so in a time shorter than the light crossing time, $t_s$, of the shell, and hence more efficiently than a classical detector. We consider this to be a quantum detection of inertial frame dragging \cite{10.1093/mnras/272.1.150,Hoefer:2014kda,PhysRevD.10.3151,KLB}. 

We note that the ability of such detectors to obtain non-local information about spacetime structure has been demonstrated in other contexts \cite{Ng:2016hzn,cong:2020crf,PhysRevD.93.044001,Lin2016,Smith_2014}. In particular, it was shown in \cite{Ng:2016hzn,cong:2020crf} that UDW detectors can distinguish between global flat Minkowkian spacetimes and local flat spacetimes inside massive shells. Our paper is an important step forward from these results, much like the transition from electrotatics to electromagnetism, or the transition from Schwarzchild to Kerr; it demonstrates for the first time that a fundamentally relativistic (non-Newtonian) effect of dragging of inertial frames, namely the existence of gravitomagnetism, can in principle be observed by a quantum detector in settings that are not classically possible.

\section{Slowly Rotating Shells}

Let us begin by describing the spacetime metric of a slowly rotating shell. The metric the shell can be written as
\begin{equation}
   \label{eq: metricO}
    \begin{split}
      ds^2_+ = -f(r) dt^2+r^2\sin^2\theta(d\phi-\frac{2 Ma}{r^3}dt)^2\\+f(r)^{-1}dr^2 +r^2 d\theta^2,
    \end{split}
\end{equation}
where $f(r)=1-2M/r$, $M$ is the mass of the shell and $a=J/M$ is the angular momentum per unit mass. The $r$-coordinate ranges from $[R,\infty)$, $R$ being the radius of the shell. To first order in $a$, the above metric agrees with the Kerr metric and satisfies the vacuum Einstein's equations. 
Inertial frame-dragging is characterized by the function  $\varpi(r) = g_{\phi t}/g_{\phi\phi}=2J/r^3$, where $J=Ma$ is the fixed total angular momentum as measured at infinity. The gradients of $\varpi(r)$ determine the precession of gyroscopes relative to the orthonormal frame of locally non-rotating observers \cite{MTW}. On the shell itself, $r=R$, and $\varpi_s = 2J/R^3$. 

For an inertial observer inside the shell (who rotates as seen from infinity) spacelike geodesics  (for example, $\phi=0, \theta=\pi/2, r = \text{constant}, t \in \mathbb{R}$) connected to fixed points at infinity rotate backwards; the shell is rotating forward (the dragging of the inertial frame becomes complete i.e., inertial observers rotate at the same angular velocity as the shell, only if the shell is at its Schwarzschild radius); the fixed stars are rotating backwards. In \cite{KLB} these effects are expressed quantitatively\footnote{In \cite{KLB} the shell is in general considered to be collapsing but the results can be immediately specialized if it is just rotating.}.

The metric  \eqref{eq: metricO} must be joined at $r=R$ to the metric
\begin{equation}
\label{eq: metricI}
\begin{split}
    ds_-^2 = -f(R) dt^2+r^2\sin^2\theta \left(d\phi-\frac{2 Ma}{R^3}dt\right)^2 \\+ d r^2+r^2 d\theta^2
        \end{split}
\end{equation}
 inside the shell, which can be seen to be flat using the coordinate transformation,
\be
\varphi = \phi - \frac{2 Ma}{R^3} t\,,
\ee
which transforms the metric~\eqref{eq: metricI} to the flat metric in standard coordinates. The stress energy tensor of the shell giving rise to the above spacetime can be found using the Israel junction condition \cite{Israel1966} and has been well-studied in the literature  \cite{poisson_2004}.

 The coordinates used in~\eqref{eq: metricO} are (spherical) Lorentzian at infinity and are naturally associated with stationary observers at infinity. All observers at fixed $(r,\theta,\varphi)$ inside the shell rotate rigidly at the rate $d\phi/dt= 2Ma/R^3$ with respect to observers at rest at infinity ($\phi=$constant). This effect is called the dragging of inertial frames, first discovered in 1918 by Thirring and Lense \cite{Lense:1918zz} and discussed in the introduction.

\section{Normalized Mode Solutions}

In order to compute the  response of the UDW detector we need to obtain the normalized mode solutions to the scalar wave equation.  For the scalar field $\Psi$ the wave equation is 
\begin{equation}
   \partial_{\mu}(\sqrt{-g}g^{\mu\nu}\partial_{\nu}\Psi) = 0\,,
\end{equation}
where $g$ is the determinant of the metric. Upon substituting in the metric~\eqref{eq: metricO} and~\eqref{eq: metricI} , this equation can be solved by separation of variables. To leading order in $a$, one can employ the usual mode expansion $\Psi_{\omega\ell m}(t,r,\theta,\phi) = \frac{1}{\sqrt{4 \pi \omega}}e^{-i\omega t}Y_{m\ell}(\theta,\phi)\psi(r)$ in spherical harmonics $Y_{m\ell}$. 
This yields a separated radial equation: 
\begin{equation}
    \label{eq: radialKG}
\frac{\alpha}{\beta \,r^2}\frac{d}{dr}\Big(\frac{\alpha}{\beta}r^2 \frac{d\psi}{dr}\Big)-\bigg(\frac{\alpha^2\ell(\ell+1)}{r^2}+\gamma +\omega^2 \bigg)\psi =0\,.
\end{equation}
The functions $\alpha$, $\beta$ and $\gamma$ are
\begin{align}
    \alpha(r) &= \begin{cases}
      \sqrt{f(R)}, & r\leq R, \\
      \sqrt{f(r)}, & r> R, \\
   \end{cases}\, \nonumber \\  \beta(r) &= \begin{cases}
      1, & r\leq R \\
      1/\sqrt{f(r)}, & r> R, \\
   \end{cases} 
\label{abc}   
   \\\gamma(r)&= \begin{cases} { 
      \frac{4Mam\omega }{R^3}-\big(\frac{2 M a m}{R^3}\big)^2}, & r\leq R, \\
       {\frac{4Mam\omega }{r^3}-\big(\frac{2 M a m}{r^3}\big)^2}, & r> R. \\
   \end{cases}\,
\nonumber    
\end{align}

For $r\leq R$, the radial equation reduces to the spherical Bessel equation, with the solution being
\be
j_{\ell}(\sqrt{b(\omega)}r)\,,\quad b(\omega)=\frac{\omega^2}{f(R)}\Big(1-\frac{2Mam}{R^3\omega}\Big)^2\,.
\ee
However, the solution outside the shell has to be determined numerically and matched to the solution on the shell.
Specifically, we impose continuity of the solution at the shell, $\psi(R) = j_{\ell}(\sqrt{b(\omega)}R)$. To find the value of $d\psi/dr|_{R^+}$, we integrate Eq.~\eqref{eq: radialKG} across the shell, obtaining the condition
\begin{equation}
\label{match}
    \bigg[\frac{\alpha(r)}{\beta(r)}\frac{d}{dr}\psi\bigg] = \left[\frac{\partial\psi}{\partial x^{\mu}} e_r^{\mu}\right] =0\,,
\end{equation}
where $e_r^{\mu}$ is the radial element of the tetrad and
the square brackets represent the difference in the value of the term across the shell. Noting from \eqref{abc} the discontinuity in  $\beta(r)$, this yields the required initial conditions $\psi(R^+)$ and $\psi'(R^+)$ for numerically solving the radial equation outside the shell.  

Finally, to normalize the solution, we will follow the scheme presented in \cite{Ng:2016hzn}.  Defining $r^{\star}$ such that $d/dr^{\star} = \frac{\alpha}{\beta}d/dr$ and  $\rho = r \psi$, the radial equation~\eqref{eq: radialKG} reads
\begin{equation}
   \frac{d^2}{dr^{\star^2}}\rho+(\omega^2-V(r))\rho=0\,,
\end{equation}
for $r>R$, where 
\begin{equation}
    V(r) = \frac{\alpha^2\ell(\ell+1)}{r^2}+\gamma+\frac{1}{r}\frac{\alpha}{\beta}\frac{d}{dr}\bigg(\frac{\alpha}{\beta}\bigg).
\end{equation}
Asymptotically, $V(r)\rightarrow 0$ as $r\rightarrow \infty$ and hence $\psi \sim \sin(\omega r^{\star})/r^{\star}$. 
Let the normalized radial solution be denoted as $\tilde{\psi}_{\omega\ell m}(r^{\star})=A_{\omega\ell m}\psi(r^{\star})$. Given any two wavefunctions $\Psi_1,\,\Psi_2$, their Klein-Gordon inner product   is 
\begin{equation}
    (\Psi_1,\Psi_2) = i\int_{\Sigma} d\sigma n^{\mu}(\Psi_1^{\star}\nabla_{\mu}\Psi_2-\Psi_2\nabla_{\mu}\Psi_1^{\star})\,,\end{equation} where $\Sigma$ is a Cauchy surface with normal $n^{\mu}$. A solution will be normalized with respect to the Klein-Gordon inner product if we choose the normalisation constant $A_{\omega\ell m}$ such that $A_{\omega\ell m}\psi(r^*)\rightarrow 2\sin(\omega r^{\star})/r^{\star}$ as $r^{\star}\rightarrow \infty$ \cite{Ng:2016hzn}.

Now that the normalised mode solutions are obtained, we are ready to compute the response of UDW detectors.

\section{UDW detector response}

A UDW detector \cite{DeWitt:1980hx,birrell_davies_1982} is a 2-level quantum mechanical system that interacts locally with a scalar quantum field as it moves along some trajectory $\sx(\tau)$ in spacetime. Letting $\Omega$ denote the energy gap of the detector and $\hat{\mu}(\tau)=e^{-i\Omega\tau}\hat\sigma^++e^{i\Omega\tau}\hat\sigma^-$ its monopole moment  (in the interaction picture), the Hamiltonian governing the detector/field  interaction is
\begin{equation}
     \hat H(\tau) = \lambda\chi(\tau)\hat\mu(\tau)\otimes\hat\Psi(\sx(\tau))\,,
\end{equation}
where $\hat{\sigma}^{\pm}$ are the ladder operators, $\tau$ is the proper time of the detector, and $\lambda$ is the dimensionless coupling constant. The duration of interaction is controlled by the switching function $\chi(\tau)$, which we will choose to be
\begin{align}
    \chi(\tau) &= \begin{cases}
      \cos^4(k \tau), &-\frac{\pi}{2 k}\leq \tau\leq\frac{\pi}{2 k}\\
       0, &\text{otherwise\,,}
   \end{cases}\label{eq:compact}
\end{align}
which has a shape similar to the Gaussian switching function $\chi_G$ \cite{cong:2020crf}
used in the static case \cite{Ng:2016hzn}, but ensures   for some $k>0$ that the interaction takes place only between $\tau\in(-\frac{\pi}{2 k},\frac{\pi}{2 k})$ We will denote the total duration of the interaction by $\Delta \tau = \pi/k$.

If the detector starts off in the ground state and interacts with the quantum vacuum via the above Hamiltonian, there may be a non-zero probability of finding the detector in its excited state after the interaction. The probability of excitation of the detector can be calculated using perturbation theory and is well-known in the literature. It is given by \cite{birrell_davies_1982,Pozas2015}
\begin{equation}
\begin{split}
    P = \lambda^2&\int_{-\infty}^{\infty}  d \tau_1\,\int_{-\infty}^{\infty}d \tau_2 \chi  (\tau_1)\chi(\tau_2) e^{-i\Omega(\tau_2-\tau_1)}\\&\times W(\sx(\tau_1),\sx(\tau_2))
    \end{split}
\end{equation}
to second order in $\lambda$, where $W(\sx(\tau_1),\sx(\tau_2))$ is the Wightman function of the field evaluated along the detector trajectory.

The field operator can be expanded in terms of the normalized field modes $\Psi_{\omega\ell m}$ of the previous section as
\begin{multline}
    \hat\psi(\sx(\tau)) =\sum_{\ell,m}\int_0^{\infty}d\omega\, \hat{a}_{\omega\ell m}\Psi_{\omega\ell m}(\sx(\tau))\\+\hat{a}^{\dagger}_{\omega\ell m}\Psi^{\dagger}_{\omega\ell m}(\sx(\tau))\,,
\end{multline}
with $\hat{a}_{\omega\ell m}$ denoting the mode annihilation operators. Let $\ket{0}$ denote the field vacuum such that $\hat{a}_{\omega\ell m}\ket{0}=0$. This corresponds to the vacuum with respect to an observer located at infinity, who is in a non-rotating frame.
The Wightman function with respect to this vacuum $W(\sx(\tau_1),\sx(\tau_2)) := 
\bra{0}{\hat\psi(\sx(\tau_2))\hat\psi(\sx(\tau_1))}\ket{0}$ is given by
\begin{equation}\label{Wight}
     W(\sx(\tau_1),\sx(\tau_2))=\sum_{\ell,m}\int_0^{\infty}d\omega \Psi^{\dagger}_{\omega\ell m}(\sx(\tau_1))\Psi_{\omega\ell m}(\sx(\tau_2))\,.
\end{equation}

From the previous section, we have seen that the normalized mode solutions are given by $\Psi_{\omega\ell m} = \frac{1}{\sqrt{4\pi\omega}}e^{-i \omega t}Y_{\ell m}(\theta,\phi)\tilde{\psi}_{\omega\ell m}(r)$. Recall that we are interested in studying how the response of the detector differs when placed respectively in a rotating shell and a stationary shell. A simple choice for the trajectory $\sx(\tau)$ of the detector is $r=r_d$, $\theta=\pi/2$, $\varphi = 0$ i.e., $\phi=\frac{2 Ma}{R^3}t$. Noting that $t=\tau/h$, where $h=\sqrt{f(R)}$, we find the \textit{response function} $\mathcal{F} = P/\lambda^2$ of the field in the form
\begin{widetext}
\begin{align}
   \nn \mathcal{F}&= \int_{-\infty}^{\infty}  d \tau_1\,\int_{-\infty}^{\infty}d \tau_2 \chi  (\tau_1)\chi(\tau_2) e^{-i\Omega(\tau_2-\tau_1)}\sum_{\ell m}\int_0^{\infty}d\omega\Psi^{\dagger}_{\omega\ell m}(\sx(\tau_1))\Psi_{\omega\ell m}(\sx(\tau_2))\\ 
   &=\sum_{\ell m}\int_0^{\infty}
\frac{d\omega}{4\pi\omega}\int_{-\infty}^{\infty}  d \tau_1\,\int_{-\infty}^{\infty}d \tau_2 \chi  (\tau_1)\chi(\tau_2) e^{-i(\Omega+\frac{\omega}{h}-\frac{2Mam}{R^3h})(\tau_2-\tau_1)}|Y_{\ell m}(\frac{\pi}{2},0)|^2|A_{\omega\ell m}|^2 |j_{\ell}(\sqrt{b(\omega)}\,  r_d)|^2\,, \nn\\
&=\sum_{\ell m}\int_0^{\infty}
\frac{d\omega}{2\omega} \left|\hat{\chi}  \left(\Omega+\frac{\omega}{h}-\frac{2Mam}{R^3 h}\right)\right|^2|A_{\omega\ell m}|^2|Y_{\ell m}(\frac{\pi}{2},0)|^2|j_{\ell}(\sqrt{b(\omega)} r_d)|^2\, ,
\label{eq: response}
\end{align}
\end{widetext}
where we switched the order of integration since the integrand is smooth and integrated over the $\tau_1$ and $\tau_2$ variables, which amounts to performing Fourier transforms 
\begin{equation}
    \hat\chi(y)=\frac{1}{\sqrt{2\pi}}\int_{-\infty}^{\infty}d\tau\chi(\tau)e^{-iy\tau} 
\end{equation}
on the switching functions, noting that  $\hat{\chi}(-y)=\hat{\chi}(y)$ for a real switching function.  

 We pause to comment that we have computed \eqref{eq: response} from the modes $\Psi_{\omega\ell m}$ assuming  \eqref{eq: radialKG} is exact.  However the metric  \eqref{eq: metricO} is a valid solution of the Einstein equations only to order $a$  {while} the leading corrections to the Wightman function
\eqref{Wight} (and thus detector response \eqref{eq: response}) are of order $a^2$.   For sufficiently small $Ma/R^2$, terms of higher order in $a$ will not significantly affect our quantitative results, 
and so we shall plot  \eqref{eq: response} in what follows.

\section{Results}

We are now ready to look at how rotation of the shell affects the response of UDW detectors. We do this by computing the expression~\eqref{eq: response} numerically, terminating the sum over $\ell$ at sufficiently large $\ell$,  chosen to give resultant errors not larger than $1\%$.
\begin{figure}[h]
    \centering
    \includegraphics[scale=0.5]{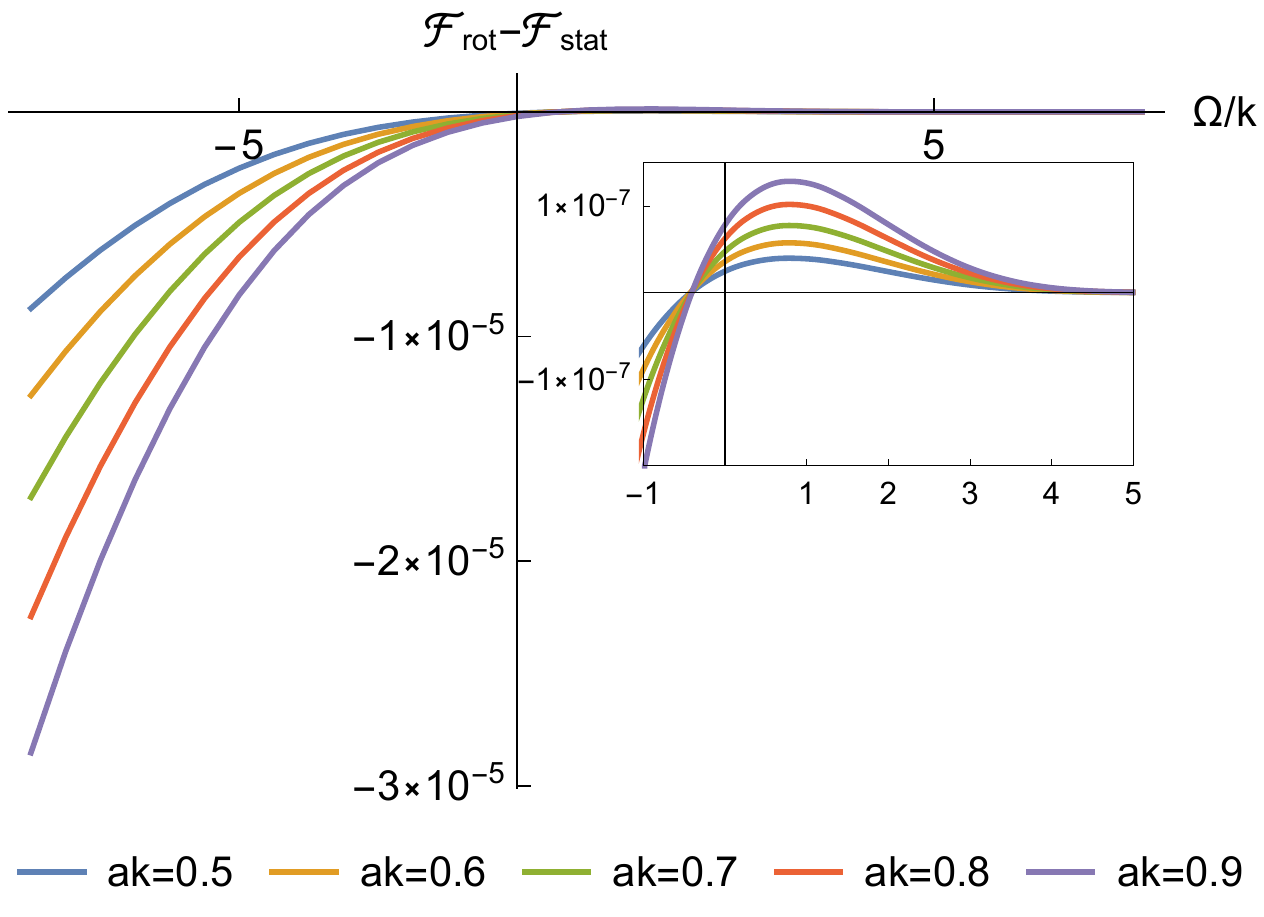}
    \caption{Detector response against $\Omega/k$. Shown here is the plot of the difference $\mathcal{F}_{rot}-\mathcal{F}_{stat}$ against $\Omega$ for different (dimensionless) rotation parameters $a k$ with $M k=1,\, R k=3\,,\,r_d k=0.5$. The inset shows a zoom-in of the plot around $\Omega/k = 0$. The difference 
    $\mathcal{F}_{rot}-\mathcal{F}_{stat}$ is small but non-zero, and is more sensitive to the rotation for negative $\Omega$.}
    \label{fig: Fvgap}
\end{figure}

Fig. \ref{fig: Fvgap} shows a  plot of $\mathcal{F}_{rot}-\mathcal{F}_{stat}\equiv \mathcal{F}_{rot}-\mathcal{F}_{rot}(a=0)$ against $\Omega$ for various (dimensionless) rotation parameters $a k$. The difference between the response of a detector placed in a slowly rotating shell $\mathcal{F}_{rot}$ and that placed in a static shell $\mathcal{F}_{stat}$, though  small, is clearly \textit{non-zero}. The difference is more pronounced when the energy gap $\Omega/k < 0$,  which physically means that the detector starts off in the excited state.  
 The rotation parameter $a$ enters the response function $\mathcal{F}$ in three positions in eq.~\eqref{eq: response}: in the Fourier transform  of the switching function, in the normalisation constant $A_{\omega\ell m}$, and in the $b(\omega)$ of the spherical Bessel function. The net effect of these is an expected increase in $|\mathcal{F}_{rot}-\mathcal{F}_{stat}|$ with $a$.

We emphasize that the interaction duration  $\Delta \tau \,k = \pi$ between the field and detector is less than  $t_s k= 2(R-r_d) k=5$, the time needed for a light signal to travel from the detector to the shell and back. This is in striking contrast to the classical case, where the fastest way a detector inside the shell (with all possible classical fields in their vacuum states) can detect the presence of rotation is by sending and waiting for a light signal to come back from the shell.

\begin{figure}
    \centering
    \includegraphics[scale=0.5]{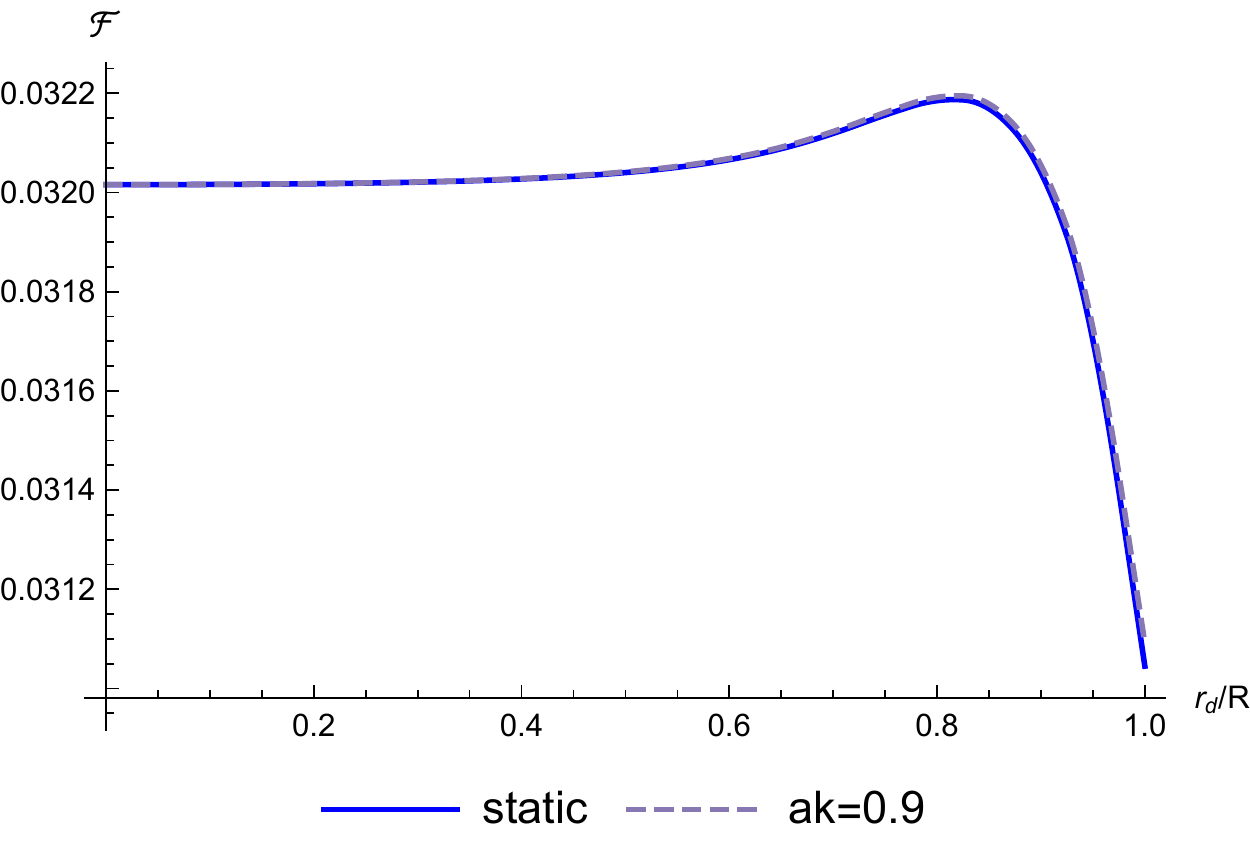}
    \includegraphics[scale=0.5]{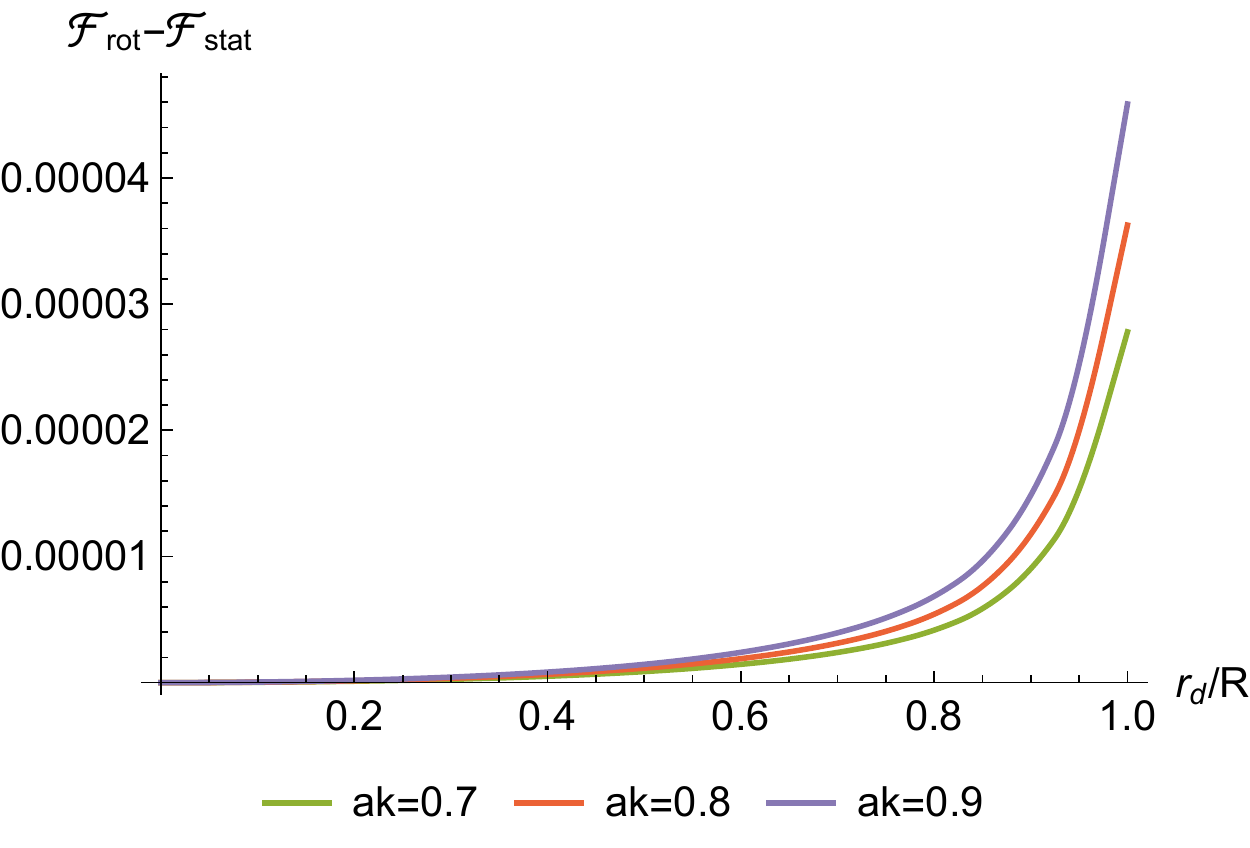}
    \caption{ Plot of $\mathcal{F}$ against $r_d/R$. These plots are obtained for $M k=1,\,$ and $R k=3$. \textbf{Above:} Plot of detector response against $r_d/R$ for static and rotating ($a k= 0.9$) shells, $\Omega/k=0.5$. \textbf{Below:} $\mathcal{F}_{rot}-\mathcal{F}_{stat}$ plots for different $a k$ settings with $\Omega / k=0.5$.}
    \label{fig: radialdistance}
\end{figure}

\begin{figure}
    \centering
    \includegraphics[scale=0.5]{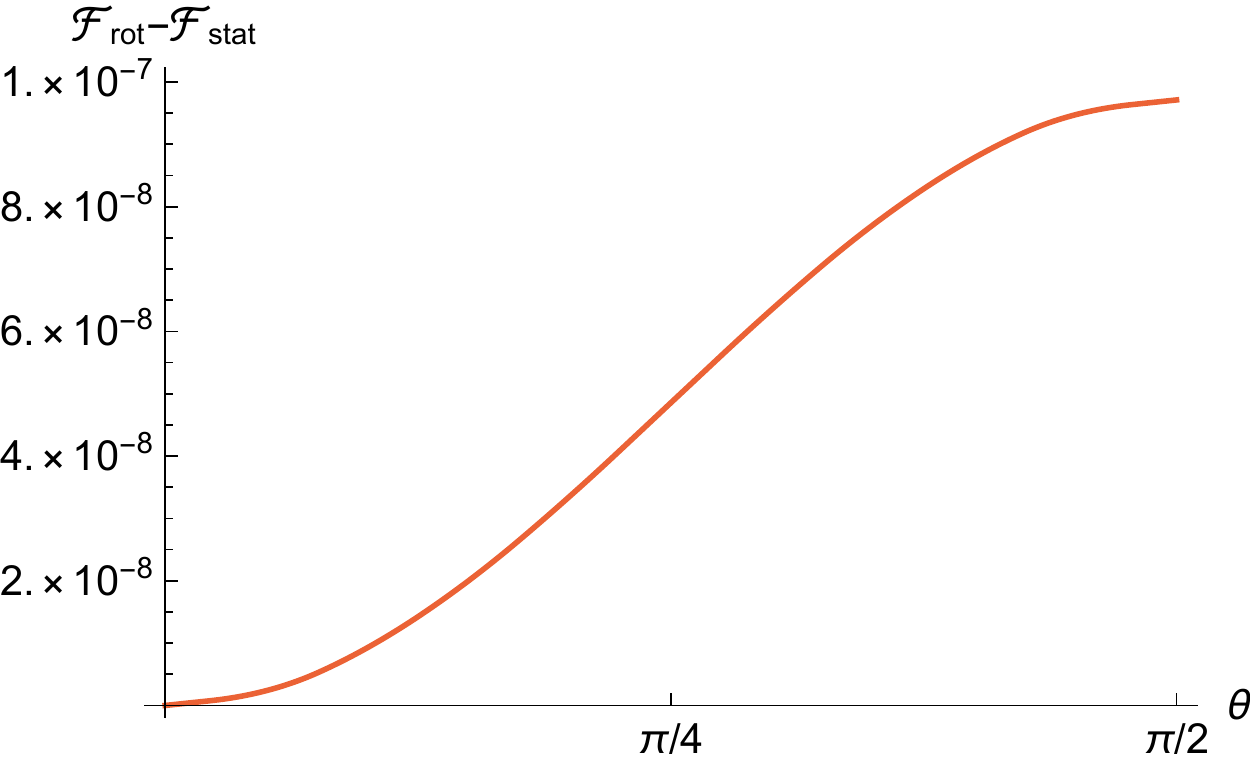}
    \caption{ Plot of $\mathcal{F}_{rot}-\mathcal{F}_{stat}$ against $\theta$ for $M k=1,\,R k=3,\,a k =0.8$ and $r_d k=0.5$.}
    \label{fig: diffvthe}
\end{figure}

In the top figure of Fig. \ref{fig: radialdistance}, we plot both $ \mathcal{F}_{rot}(ak=0.9)$ and $\mathcal{F}_{stat}$ against the detector location $r_d/R$. The responses peak at some intermediate $r_d$, in agreement with the results of ref.\cite{cong:2020crf}. From the bottom figure, we see that the detector response 
increases by more than an order of magnitude as compared to Fig.~\ref{fig: Fvgap} as $r_d/R\to 1$.   We find that the shape of the curves in Fig~\ref{fig: Fvgap} remains qualitatively the same as $r_d/R$ increases, though the interaction duration is eventually no longer less than the light crossing time.  A detector
placed at the origin $r_d=0$ cannot distinguish between a rotating and a static shell. We can  understand this explicitly by noting that the rotation parameter $a$ appears in the radial equation~\eqref{eq: radialKG} through the term $\gamma$, where it is multiplied with the azimuthal number $m$. Hence, it has only nontrivial effects when $m\neq 0$. However since $\theta=0$ along the axis of rotation and $Y_{\ell m}(0,0)$ is non-zero only when $m=0$, the mode solutions and hence the response function are insensitive to  effects of rotation along this axis. As another illustration of this, we plot in Fig. \ref{fig: diffvthe} $\mathcal{F}_{rot}-\mathcal{F}_{stat}$ against $\theta$, the angle measured from the rotation axis. From this, we see that the sensitivity to rotation of detectors placed at the same $r_d$ increases monotonically as $\theta$ increases from $0$ to $\pi/2$.

\section{Conclusions}

Classically, the physical effect of a slowly rotating shell is the dragging of inertial frames. We have shown that this effect can be discerned from local measurements of a quantum particle detector
inside the shell, on timescales much shorter than the light travel time from the detector to the edge of the shell and back. 

We note that the gravitational effects inside a rotating material shell are analogous to the electromagnetic effects inside a rotating charged shell; but there are also fundamental differences. For a rotating charged shell, a dipolar magnetic field will be formed inside. Such a field can be observed without the need of quantum detectors, for example as the Larmor precession of charged particles. 

By solving the scalar field equation numerically, we have obtained the response function of the detector and seen how it depends on the rotation parameter $a$.
 Corrections to the metric  \eqref{eq: metricO} to higher orders in $a$ will quantitatively modify  \eqref{eq: response} but will not qualitatively affect our results. Alternatively, we can regard \eqref{eq: metricO}   as a `kinematic spacetime'  that could be employed in analogue gravity laboratory simulations, in which case our results would hold exactly.  Whether or not such effects can be directly detected remains a challenge for future experiments.

 \section*{Acknowledgments}
 \label{sc:acknowledgements}
 We would like to thank Keith Ng for helpful discussions on this work and Markus King for a helpful correspondence. This work was supported in part by the Natural Sciences and Engineering Research Council of Canada and by the Perimeter Institute. J.B. thanks the kind hospitality of the Perimeter Institute, Waterloo, where this work started and the Albert-Einstein Institute, Golm, where it continued. J.B. also acknowledges the support from the Grant Agency of the Czech Republic, Grant No. GA\u CR 19-01850S. Research at Perimeter Institute is supported in part by the Government of Canada through the Department of Innovation, Science and Economic Development Canada and by the Province of Ontario through the Ministry of Colleges and Universities.
 
\bibliography{UDWshellv33}

\end{document}